\documentclass[aps,prl,twocolumn,showpacs,amsmath,amssymb,superscriptaddress]{revtex4}
\usepackage{graphicx}
\usepackage{amssymb}
\usepackage{natbib}
\usepackage{color}

\newcommand{\beq}{\begin{eqnarray}}
\newcommand{\eeq}{\end{eqnarray}}

\begin{document}
\title{Strain-induced spin-nematic state and nematic susceptibility arising from $2\times2$ Fe clusters in KFe$_{0.8}$Ag$_{1.2}$Te$_2$}
\author{Yu Song}
\email{yusong@berkeley.edu}
\affiliation{Department of Physics, University of California, Berkeley, California 94720, USA}
\affiliation{Materials Sciences Division, Lawrence Berkeley National Laboratory, Berkeley, California 94720, USA}
\author{Dongsheng Yuan}
\affiliation{Materials Sciences Division, Lawrence Berkeley National Laboratory, Berkeley, California 94720, USA}
\author{Xingye Lu}
\affiliation{Center for Advanced Quantum Studies and Department of Physics, Beijing Normal University, Beijing 100875, China}
\author{Zhijun Xu}
\affiliation{NIST Center for Neutron Research,National Institute of Standards and Technology, Gaithersburg MD 20899, USA}
\affiliation{Department of Materials Science and Engineering,University of Maryland, College Park, MD 20742, USA}
\author{Edith Bourret-Courchesne}
\affiliation{Materials Sciences Division, Lawrence Berkeley National Laboratory, Berkeley, California 94720, USA}
\author{Robert J. Birgeneau}
\affiliation{Department of Physics, University of California, Berkeley, California 94720, USA}
\affiliation{Materials Sciences Division, Lawrence Berkeley National Laboratory, Berkeley, California 94720, USA}
\affiliation{Department of Materials Science and Engineering, University of California, Berkeley, California 94720, USA}

\begin{abstract}	
Spin nematics break spin-rotational symmetry while maintaining time-reversal symmetry, analogous to liquid crystal nematics that break spatial rotational symmetry while maintaining translational symmetry. Although several candidate spin nematics have been proposed, the identification and characterization of such a state remain challenging because the spin-nematic order parameter does not couple directly to experimental probes. 
KFe$_{0.8}$Ag$_{1.2}$Te$_2$ (K$_5$Fe$_4$Ag$_6$Te$_{10}$, KFAT) is a local-moment magnet consisting of well-separated 2$\times$2 Fe clusters, and in its ground state the clusters order magnetically, breaking both spin-rotational and time-reversal symmetries. 
Using uniform magnetic susceptibility and neutron scattering measurements we find a small strain induces sizable spin anisotropy in the paramagnetic state of KFAT,
manifestly breaking spin-rotational symmetry while retaining time-reversal symmetry, resulting in a strain-induced spin-nematic state in which the $2\times2$ clusters act as the spin analogue of molecules in a liquid crystal nematic.  
The strain-induced spin anisotropy in KFAT allows us to probe its nematic susceptibility,
revealing a divergent-like increase upon cooling, indicating the ordered ground state is driven by a  spin-orbital entangled nematic order parameter.
  
\end{abstract}

\pacs{74.25.Ha, 74.70.-b, 78.70.Nx}

\maketitle

Quantum materials often adopt electronic ground states that break rotational-symmetry of their underlying crystal structures, analogous to liquid crystal nematics \cite{EFradkin2010}. Such electronic nematic states are immensely interesting in their own right, and have additionally garnered recent attention due to their ubiquity near unconventional superconducting states \cite{YAndo2002,SMukhopadhyay2019,JHChu1,HHKuo2016,FRonning}. 
One form of such a state is the spin nematic, which maintains time-reversal symmetry but breaks spin-rotational symmetry through an even-order spin order parameter such as $\langle (S^x)^2\rangle-\langle (S^y)^2\rangle\neq0$ \cite{MBlume,AFAndreev,KPenc}, in contrast to conventional magnetic order such as $\langle S^x\rangle-\langle S^y\rangle\neq0$ that breaks both symmetries. Examples of the spin-nematic state have been proposed in frustrated magnets \cite{KPenc,YKohama}, rare-earth magnets \cite{ROkazaki,EWRosenberg} and iron-based superconductors (IBS) \cite{CFang2008,CXu2008,RYu2015}. However, because the order parameters for proposed spin-nematic states do not couple directly to experimental probes, a direct detection of the spin-nematic state remains elusive.  

The parent compounds of iron-based superconductors (PC-IBS) are bad metals made up of stacked Fe square lattices forming a tetragonal structure, and exhibit an orthorhombic distortion below $T_{\rm S}$ as well as a stripe-type magnetic ordering below $T_{\rm N}$ ($T_{\rm N}\leq T_{\rm S}$) \cite{GStewart_RMP,PDai_RMP}. In the paramagnetic tetragonal state above $T_{\rm S,N}$, a sizable resistivity anisotropy characteristic of an electronic nematic state ($\eta$) can be induced by a small strain ($\epsilon$) \cite{JHChu1,MTanatar2016}, and divergence of the nematic susceptibility ($d\eta/d\epsilon$) demonstrates the phase transitions below $T_{\rm S,N}$ are driven by an electronic nematic order parameter that couples to the lattice \cite{JHChu2}. The electronic nematic order parameter may be associated with ferro-orbital ordering \cite{WLv2009,CCLee2009} or a spin-nematic state \cite{CFang2008,CXu2008}, and in the latter anisotropy in spin correlations between the two Fe-Fe directions breaks spin-rotational symmetry while retaining time-reversal symmetry 
\cite{CFang2008,CXu2008,RMFernandes2012_1,RMFernandes2012_2,SLiang2013}. Experimental evidence for spin-nematicity in these materials include scaling of the shear modulus and spin-lattice relaxation rate \cite{RMFernandes2013} and nematic spin correlations observed in neutron scattering \cite{XLu2014,YSong2015}, however the metallic nature of these materials means the shape \cite{HZhai2009,YGallais2013} and orbital content \cite{MYi2011} of the Fermi surface may play major roles accounting for these observations. In addition, while anisotropy of the uniform magnetic susceptibility is anticipated to scale with the resistivity anisotropy in the spin-nematic scenario \cite{RMFernandes2012_1,RMFernandes2012_2}, experimentally such anisotropies have not been detected \cite{MHe2017,MHe2018}.    

KFAT is structurally similar to PC-IBS, but rather than a metal it is a semiconductor with localized magnetism \cite{HLei2011,RAng2013}, and instead of planes of Fe square lattices it consists of 2$\times$2 Fe blocks well separated by nonmagnetic Ag atoms [Fig. 1(b)] \cite{YSong2019}. Nonetheless, diffraction measurements show that it exhibits simultaneous stripe-type magnetic and structural phase transitions below $T_{\rm S,N}\approx35$~K, similar to the PC-IBS such as BaFe$_2$As$_2$ [Fig. 1(a), (b)] \cite{QHuang,YSong2019}. 
This suggests that similar to IBS \cite{JHChu2}, an electronic nematic order parameter coupled to the lattice may be present in KFAT; and given its semiconducting nature, such an electronic nematic order parameter should arise from localized spin and orbital degrees of freedom, allowing the interplay between different orders in IBS to be probed in the strong-coupling limit \cite{TYildirim2008,QMSi2008,FKruger2009} within $2\times2$ clusters.     

In this Letter, we demonstrate that a small strain induces a sizable uniform magnetic susceptibility anisotropy without conventional magnetic order in KFAT. The observed anisotropy demonstrates the presence of an emergent Ising degree of freedom associated with the $2\times2$ spin clusters, and their collective ordering under strain realizes a strain-induced spin-nematic state, with the clusters acting as the spin analogue of rod-like molecules in liquid crystal nematics. The nematic susceptibility of the spin anisotropy increases strongly upon cooling towards $T_{\rm S,N}$, indicating that spin-nematicity drives the ordered ground state. The strong coupling between lattice and spin-nematicity results from an underlying spin-orbital entangled state, allowing for a novel form of elastomagnetic effect, in which strain induces anisotropic uniform magnetic susceptibility in a paramagnetic state.   

Single crystals of KFAT are grown from a stoichiometric mixture of elemental materials using a modified Bridgeman method \cite{HLei2011}. Magnetization measurements were carried out using a Quantum Design MPMS3 SQUID under a field of 5~T. The magnetization in KFAT is linear with field up to at least 5~T \cite{HLei2011}.
Neutron scattering measurements were carried out using the BT-7 triple-axis spectrometer at the NIST Center for Neutron Research (NCNR), aligned in the $[H,K,0]$ plane. Strain is applied by gluing $\sim$0.2~mm thick platelet samples on a glass-fiber-reinforced plastic (GFRP) substrate embedded with unidirectional glass fibers, using STYCAST 1266 two part epoxy [Fig. 2(a)] \cite{MHe2017}. 
While KFAT exhibits a $\sqrt{5}\times\sqrt{5}$ superstructure, its physics is dominated by interactions within the 2$\times$2 Fe clusters; we therefore use the $I4$/$mmm$ unit cell appropriate for tetragonal BaFe$_2$As$_2$ to describe our results [Fig. 1(c)], in this notation the Fe-Fe bonds are along $(110)$/$(1\bar{1}0)$ directions. 

\begin{figure}[t]
	\includegraphics[scale=.50]{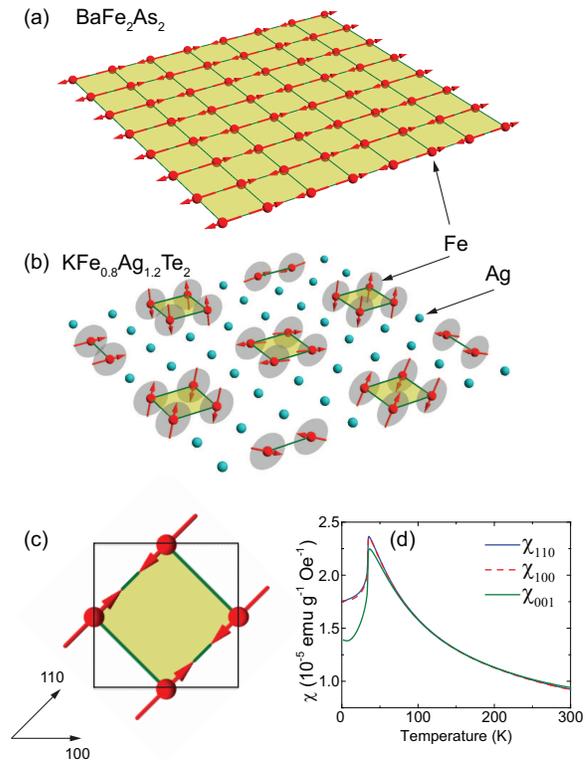}
	\caption{
		(Color online) (a) The magnetic structure of BaFe$_2$As$_2$, a prototypical PC-IBS. (2) The crystal and magnetic structures of KFAT. The shaded circles represent the magnetic easy plane. (c) The magnetic structure of KFAT projected into the $ab$-plane. The solid box is the $I4$/$mmm$ unit cell used in this work. (d) The magnetic susceptibility of freestanding KFAT measured along three high-symmetry directions. $\chi_{110}$ and $\chi_{100}$ almost overlap.    
	}
\end{figure}

Below $T_{\rm S,N}$, a collinear stripe-type spin configuration forms inside each 2$\times$2 cluster of KFAT, with antiferromagnetic (AF) alignment along one Fe-Fe direction and ferromagnetic (FM) alignment along the other. 
The collinear spin direction rotates in an easy-plane from cluster to cluster, approximately spanned by the AF Fe-Fe direction and the $c$-axis [Fig. 1(b)] \cite{YSong2019}. When projected into the $ab$-plane, the magnetic structure within each cluster resembles the PC-IBS [Fig. 1(a), (c)], with the relative easy-axis along the AF Fe-Fe direction. Magnetic susceptibilities of freestanding KFAT along three high-symmetry directions are shown in Fig. 1(d), in good agreement with previous report \cite{HLei2011}. The Curie-Weiss behavior of the magnetic susceptibilities above $T_{\rm S,N}$ with an effective moment $\mu_{\rm eff}\approx2.8 \mu_{\rm B}$/Fe ($S\approx1$) \cite{SI} points to local-moment magnetism, consistent with its semiconducting transport \cite{HLei2011}. The larger drop of the magnetic susceptibility along the $c$-axis below $T_{\rm S,N}$ is consistent with it spanning the easy-plane; however due to the formation of twin domains below $T_{\rm S,N}$, the magnetic susceptibility appear isotropic in the $ab$-plane for the freestanding sample.   

To probe the intrinsic magnetic susceptibility of KFAT, it is necessary to detwin the sample. The structural distortion accompanying the magnetic order in KFAT differentiates the lattice spacings along $(110)$/$(1\bar{1}0)$,  with the AF (FM) Fe-Fe direction elongated (contracted) below $T_{\rm S,N}$; it is therefore possible to detwin the sample by applying strain, similar to the PC-IBS \cite{CDhital2012,YSong2013,XLu2016}. By gluing the sample onto a GFRP substrate, strain is applied due to anisotropic thermal contraction of the substrate \cite{MHe2017}. Upon cooling, the direction parallel (perpendicular) to the unidirectional fibers will be the relatively longer (shorter) axis [Fig.~2(a)], and since the AF (FM) aligned Fe-Fe direction elongates (contracts) below $T_{\rm S,N}$, the majority domain has the AF (FM) Fe-Fe direction parallel (perpendicular) to the fibers. 

Using this setup, we measured the magnetizations parallel ($M_{\rm para}$) and perpendicular ($M_{\rm perp}$) to the fibers with strain along $(110)$/$(1\bar{1}0)$, revealing a clear anisotropy 
[Fig.~2(c)]. The magnetic susceptibilities $\chi_{\rm para}$ and $\chi_{\rm perp}$ are obtained by subtracting the magnetization due to the substrate ($M_{\rm GFRP}$) and dividing by the applied field [Fig.~2(d)]. Upon cooling below $T_{\rm S,N}$, $\chi_{\rm para}$ exhibits a more prominent drop compared to $\chi_{\rm perp}$, suggesting that for the two in-plane directions, spins order along the AF Fe-Fe direction (which dominates $\chi_{\rm para}$).
Combined with $\chi_{001}$ in Fig.~1(d), these measurements demonstrate that below $T_{\rm S,N}$, KFAT exhibits a magnetic easy-plane spanned by the AF Fe-Fe direction and the $c$-axis, consistent with previous diffraction results \cite{YSong2019}. 

\begin{figure}[t]
	\includegraphics[scale=.5]{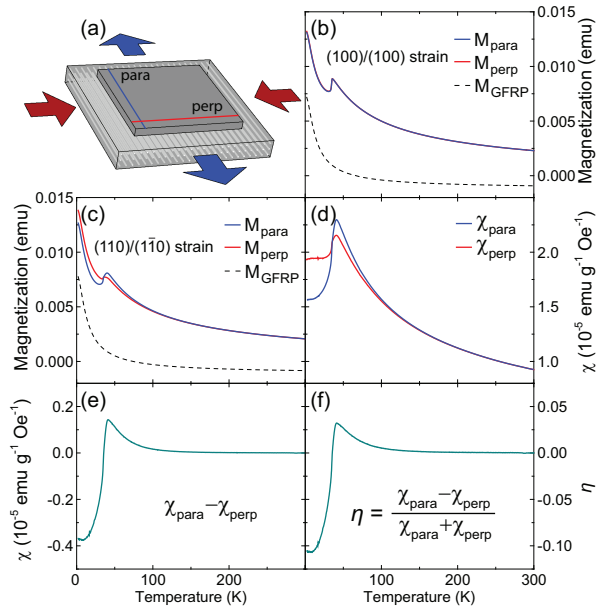}
	\caption{ 
		(Color online) (a) Schematic of our experimental setup with the sample glued onto a GFRP substrate. Upon cooling, the direction parallel (perpendicular) to the glass fibers become relatively longer (shorter), applying strain to the sample (arrows). This setup is used for magnetization measurements parallel and perpendicular to the fibers, for (b) $(100)$/$(010)$ strain and (c) $(110)$/$(1\bar{1}0)$ strain. (d) Magnetic susceptibilities parallel ($\chi_{\rm para}$) and perpendicular ($\chi_{\rm perp}$) to the fibers. (e) $\chi_{\rm para}-\chi_{\rm perp}$. (f) $\eta=(\chi_{\rm para}-\chi_{\rm perp})/(\chi_{\rm para}+\chi_{\rm perp})$.
	}
\end{figure}

Unexpectedly, we also observed sizable magnetic susceptibility anisotropy over an extended temperature range for $T\gtrsim T_{\rm S,N}$ [Fig.~2(d)], in contrast to similar experiments on BaFe$_2$As$_2$ above $T_{\rm S,N}$ and FeSe above $T_{\rm S}$ \cite{MHe2017,MHe2018}. To quantify the observed anisotropy, the difference $\chi_{\rm para}-\chi_{\rm perp}$ and the dimensionless anisotropy $\eta=(\chi_{\rm para}-\chi_{\rm perp})/(\chi_{\rm para}+\chi_{\rm perp})$ are respectively shown in Fig.~2(e) and 2(f). As can be seen, the sign of the anisotropy changes from $\chi_{\rm para}<\chi_{\rm perp}$ for $T\lesssim T_{\rm S,N}$ to $\chi_{\rm para}>\chi_{\rm perp}$ for $T\gtrsim T_{\rm S,N}$. As discussed above, $\chi_{\rm para}<\chi_{\rm perp}$ below $T_{\rm S,N}$ is because $\chi_{\rm para}$ probes the magnetic easy-axis (AF Fe-Fe direction), which has a small magnetic susceptibility in the ordered state; the reversed anisotropy for $T\gtrsim T_{\rm S,N}$ with $\chi_{\rm para}>\chi_{\rm perp}$ instead indicates a paramagnetic state, in which a larger magnetic susceptibility is expected along the easy-axis. Our results therefore show that a small strain induces a sizable spin-rotational symmetry-breaking without conventional magnetic order, realizing a strain-induced spin-nematic state. 

Similar measurements were carried out for $(100)$/$(010)$ strain [Fig.~2(b)]; we find $M_{\rm para}$ and $M_{\rm perp}$ to be essentially identical for all temperatures (difference less than 0.05\% of their values for $T\gtrsim T_{\rm S,N}$), in contrast to $(110)$/$(1\bar{1}0)$ strain [Fig.~2(c)]. This demonstrates that strain along Fe-Te directions does not detwin the sample below $T_{\rm S,N}$, and more importantly, the strain-induced spin anistropy in the paramagnetic state exhibits a prominent Ising character (at least $\approx100$ times larger for $(110)$/$(1\bar{1}0)$ strain). Such a prominent Ising-type response is similar to resistivity anisotropies in Sr$_3$Ru$_2$O$_7$ (field-induced) \cite{RABorzi2007} and BaFe$_2$As$_2$ (strain-induced) \cite{ZLiu2016}, but contrasts with CeRhIn$_5$ (field-induced) \cite{FRonning} and heavily electron- or hole-doped IBS (strain-induced) \cite{HHKuo2013,ZLiu2016,JLi2016,XLiu2019,KIshida2018}, in which the Ising character of the response is weaker and even XY-like.

\begin{figure}[t]
	\includegraphics[scale=.5]{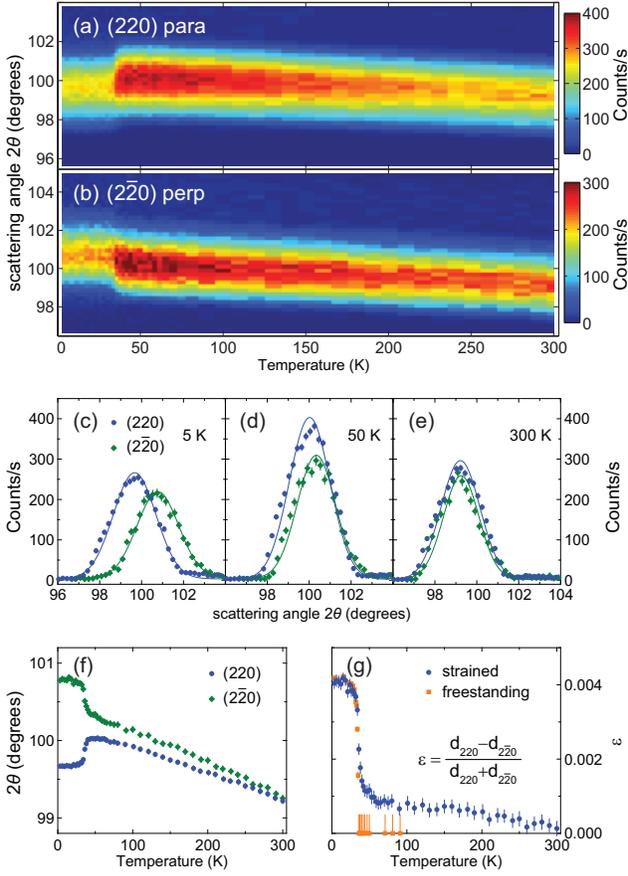}
	\caption{ 
		(Color online) Pseudo-color plots of longitudinal scans for (a) $(220)$ and (b) $(2\bar{2}0)$ Bragg peaks as a function of temperature. Representative scans are compared for (c) $T=5$~K, (d) $T=50$~K and (e) $T=300$~K. (f) Scattering angles $2\theta$ for longitudinal scans in (a) and (b), obtained from fits to Gaussian peaks. (g) The lattice anisotropy $\epsilon$ (strain), for a sample under strain and a freestanding sample. The solid lines in (c), (d) and (e) are the results of fits to Gaussian peaks.
	}
\end{figure}

Having shown that strain induces a sizable spin anisotropy in KFAT, we elucidate its origin with neutron scattering using the same strain setup. For a sample with $(110)$/$(1\bar{1}0)$ strain, we monitored its $(220)$ (parallel to fibers) and $(2\bar{2}0)$ (perpendicular to fibers) Bragg peaks as a function of temperature, with results shown in Figs.~3(a) and (b) and compared for selected temperatures in Figs.~3(c)-(e). The scattering angles for the two Bragg peaks are obtained through Gaussian fits [Fig.~3(f)]. Upon cooling, the scattering angles for both peaks increase monotonically due to thermal contraction down to $T\approx T_{\rm S,N}$, below which $(220)$ moves to lower scattering angles and $(2\bar{2}0)$ moves to higher scattering angles. The disparate responses are due to detwinning of the sample, in contrast to a twinned sample in which splittings (or broadenings when resolution is insufficient) of both $(220)$ and $(2\bar{2}0)$ are observed \cite{YSong2019}.
From the scattering angles we obtained the corresponding lattice spacings $d_{220}$ and $d_{2\bar{2}0}$, and extracted the temperature dependence of the strain $\epsilon=(d_{220}-d_{2\bar{2}0})/(d_{220}+d_{2\bar{2}0})$ [Fig.~3(g)], which is a dimensionless measure of the lattice anisotropy. Compared to a freestanding sample, the similar values of $\epsilon$ (orthorhombicity for $T<T_{\rm S,N}$) indicates that our strain setup mainly acts to detwin the sample below $T_{\rm S,N}$, while for $T\gtrsim T_{\rm S,N}$ the structural transition is smeared out and a non-zero strain is induced up to room temperature.

\begin{figure}[t]
	\includegraphics[scale=.5]{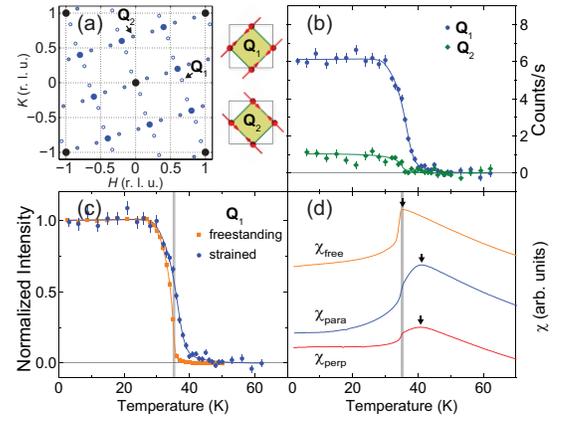}
	\caption{ 
		(Color online) (a) Allowed structural and magnetic peaks in the $[H,K,0]$ scattering plane of KFAT. Superstructure peaks occur due to the $\sqrt{5}\times\sqrt{5}$ ordering between Fe and Ag, and incommensurate magnetic peaks straddle allowed structural Bragg peaks. ${\bf Q}_1$ and ${\bf Q}_2$ are two magnetic Bragg peaks related by a $90^{\circ}$ rotation, corresponding to magnetic domains with AF spins along $(110)$ and $(1\bar{1}0)$, respectively. Only peaks associated with one of two $\sqrt{5}\times\sqrt{5}$ superstructures are shown \cite{YSong2019}. (b) Temperature dependence of ${\bf Q}_1$ and ${\bf Q}_2$ magnetic Bragg peaks measured with $(110)$ parallel to fibers. (c) Temperature dependence of the ${\bf Q}_1$ magnetic peak in a freestanding and a strained sample compared, after normalizing by the intensity at low temperatures. (d) Uniform magnetic susceptibilities of a freestanding and a strained sample compared near $T_{\rm S,N}$.  
	}
\end{figure}

To clarify the effect of strain on the magnetic order in KFAT, we studied two magnetic Bragg peaks related by a 90$^{\circ}$ rotation, with domains associated with ${\bf Q}_1$ (${\bf Q}_2$) having the longer AF (shorter FM) Fe-Fe bonds parallel to the fibers [Fig.~4(a)]. The temperature dependence of these two magnetic Bragg peaks are shown in Fig.~4(b), from their ratio a detwinning ratio of $\approx$6:1 is obtained, providing additional evidence that applied strain effectively detwins the sample, with the majority domain having its longer AF Fe-Fe bonds parallel to the fibers. By comparing the temperature dependence of the ${\bf Q}_1$ Bragg peak in a strained sample and a freestanding sample [Fig.~4(c)], we find that strain enhances the onset temperature of the magnetic transition by several Kelvins, similar to the PC-IBS \cite{CDhital2012,YSong2013,XLu2016}. The increase of onset temperature of the magnetic transition under strain can also be seen from our magnetic susceptibility measurements, with results in Figs. 1(d) and 2(d) magnified and compared in Fig.~4(d).
We note that while magnetic order onsets below $\approx$50~K under strain, the uniform magnetic susceptibility anisotropy $\chi_{\rm para}>\chi_{\rm perp}$ extends to much higher temperatures.

An important question to address in KFAT is whether its structural transition results from a lattice instability or an underlying electronic nematic order parameter. 
To resolve this issue, consider the free energy
\begin{equation*} 
F=\frac{a}{2}\phi^2+\frac{b}{4}\phi^4+\frac{c}{2}\epsilon^2+\frac{d}{4}\epsilon^4-\lambda\phi\epsilon,
\end{equation*}
where $\phi$ is an electronic nematic order parameter and $\epsilon$ is the lattice distortion, with $\phi$ and $\epsilon$ coupled through $\lambda$ \cite{JHChu2}. As previously shown, the strain-nematic susceptibility $d\phi/d\epsilon$ will exhibit a divergence only if the structural phase transition is driven by an instability in $\phi$ \cite{JHChu2}. In the case of KFAT, we have obtained the dimensionless spin [$\eta$, Fig.~2(f)] and lattice [$\epsilon$, Fig.~3(g)] anisotropies under strain. The measured spin anisotropy $\eta\propto\phi$, and within linear response $d\phi/d\epsilon\propto d\eta/d\epsilon=\eta/\epsilon$. From our measurements, $\eta/\epsilon$ for KFAT [Fig.~5(a)] exhibits a clear divergence-like increase upon cooling towards $T_{\rm S,N}$, indicating the coupled phase transitions at $T_{\rm S,N}$ are driven by an instability in $\phi$ \cite{SI}. The nematic susceptibility for $T\geq50$~K can be described by a Curie-Weiss form $\eta/\epsilon=a/(T-T^*)+b$ with $T^*=29(3)$~K. Since both $\eta$ and $\epsilon$ are dimensionless anisotropies, $\eta/\epsilon$ provides a quantitative comparison between the two. Cooling towards $T_{\rm S,N}$, $\eta/\epsilon\approx30$ at $T\approx50$ K, the much large spin anisotropy indicates an inherent tendency towards breaking spin-rotational symmetry; although in the absence of strain, time-reversal symmetry simultaneously breaks and results in conventional magnetic order.

Compared to our findings in KFAT, similar magnetic susceptibility anisotropies were not detected in strained BaFe$_2$As$_2$ and FeSe \cite{MHe2017,MHe2018}, likely due to intrinsically smaller magnetic susceptibilities in IBS \cite{XFWang2009,JYang2010}, and development of the spin-nematic order parameter at the stripe-type ordering vector is accompanied by an increase in the magnetic correlation length \cite{WZhang2016} that reduces effects of the former in the uniform limit. In contrast, the local-moment nature KFAT results in a larger magnetic susceptibility, and the $2\times2$ clusters limit effects of the magnetic correlation length, making KFAT more amenable for the detection of magnetic susceptibility anisotropy.

\begin{figure}[t]
	\includegraphics[scale=.45]{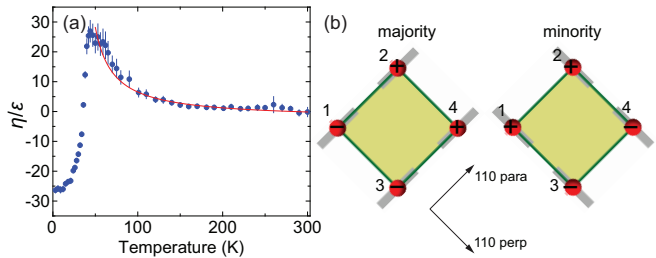}
	\caption{ 
		(Color online) (a) $\eta/\epsilon$. The red line is a Curie-Weiss fit for $T\geq50$~K, with the Weiss temperature $T^*=29(3)$~K. We note $T^*$ also has a dependence on the temperature range used in fitting. (b) Two stripe-type configurations of $2\times2$ Fe clusters with corresponding local spin anisotropies (easy-axis along shaded slabs). The local spin anisotropy is present in the ordered state, as well as in the paramagnetic state when the clusters fluctuate between the two configurations. In the strained paramagnetic state, the clusters fluctuate in one of the configurations more than the other, resulting in a spin-nematic state with macroscopic spin anisotropy. 
	}
\end{figure}
Given the stripe-type ordered ground state of KFAT below $T_{\rm S,N}$, a natural candidate to account for the spin-nematicity is an Ising degree of freedom emerging from the  
$2\times2$ spin clusters: $\Phi=\langle {\bf S}_1\cdot {\bf S}_3\rangle+\langle {\bf S}_2\cdot {\bf S}_4\rangle-\langle {\bf S}_1\cdot {\bf S}_2\rangle-\langle {\bf S}_3\cdot{\bf S}_4\rangle$ [Fig.~5(b)], a vestigial order of magnetic order \cite{RMFernandes2019} similar to the spin-nematic state proposed for the IBS \cite{CFang2008}.  
Our observation of strain having a strong effect on the spin anisotropy immediately points to a strong coupling between the spin and orbital degrees of freedom. One form of such coupling is of the Kugel-Khomskii type \cite{KK,Oles2012,FKruger2009}, in which strain-induced lattice anisotropy results in both ordering between $d_{xz}$/$d_{yz}$ orbitals and $\Phi\neq0$, and the shape of $d_{xz}$/$d_{yz}$ orbitals locks the AF (FM) Fe-Fe direction with the elongated (contracted) crystal axis due to bond-dependent exchange couplings. 
Another essential ingredient for our observation is the presence of a strong local spin anisotropy [easy-axis along grey slabs in Fig.~5(b)], present in the magnetically ordered ground state and persisting well above $T_{\rm S,N}$ when the clusters are fluctuating independently between two dynamic configurations with $\Phi>0$ and $\Phi<0$. We note while the magnetic order in the PC-IBS is collinear with spins along the AF Fe-Fe direction, the $c$-axis polarized spin waves are lowest in energy \cite{NQureshi2012,YSong2013_2}. This indicates a common hierarchy of spin anisotropy energies $\Delta_{\rm AF}\leq\Delta_{c}<\Delta_{\rm FM}$ in PC-IBS and KFAT. 
These two effects in combination give rise to an spin-orbital entangled state, resulting in a unusually large coupling between spin and the lattice, allowing the spin-nematic state to directly manifest through uniform magnetic susceptibility measurements under strain. 

In conclusion, we have demonstrated that the $2\times2$ spin clusters in KFAT gives rise to an emergent spin-orbital entangled Ising degree of freedom, analogous to molecules in liquid crystal nematics. This emergent degree of freedom drives the coupled phase transitions and allows for a sizable spin anisotropy to be induced by a small strain, with the spin clusters collectively forming a spin nematic.

We acknowledge helpful discussions with Dung-hai Lee and Mingquan He. The work at Lawrence Berkeley National Laboratory was supported by the Office of Science, Office of BES, Materials Sciences and Engineering Division, of the US DOE under contract No. DE-AC02-05-CH11231 within the Quantum Materials Program (KC2202). The work at Beijing Normal University was supported by the National Natural Science Foundation of China under Grant No. 11734002. The identification of any commercial product or trade name does not imply endorsement or recommendation by the National Institute of Standards and Technology.


\begin{thebibliography}{}
\bibitem{EFradkin2010} Eduardo Fradkin, Steven A. Kivelson, Michael J. Lawler, James P. Eisenstein, and Andrew P. Mackenzie, Annu. Rev. Condens. Matter Phys. {\bf 1}, 153-178 (2010).  	
	
\bibitem{YAndo2002} Yoichi Ando, Kouji Segawa, Seiki Komiya, and A. N. Lavrov, Phys. Rev. Lett. {\bf 88}, 137005 (2002).

\bibitem{SMukhopadhyay2019} Sourin Mukhopadhyay, Rahul Sharma, Chung Koo Kim, Stephen D. Edkins, Mohammad H. Hamidian, Hiroshi Eisaki, Shin-ichi Uchida, Eun-Ah Kim, Michael J. Lawler, Andrew P. Mackenzie, J. C. S\'{e}amus Davis, and Kazuhiro Fujita, Proc. Natl. Acad. Sci. U.S.A. {\bf 116}, 13249-13254 (2019). 

\bibitem{JHChu1} Jiun-Haw Chu, James G. Analytis, Kristiaan De Greve, Peter L. McMahon, Zahirul Islam, Yoshihisa Yamamoto, and Ian R. Fisher, Science {\bf 329}, 824-826 (2010). 

\bibitem{HHKuo2016} Hsueh-Hui Kuo, Jiun-Haw Chu, Johanna C. Palmstrom, Steven A. Kivelson, Ian R. Fisher, Science {\bf 352}, 958-962 (2016).

\bibitem{FRonning} F. Ronning, T. Helm, K. R. Shirer, M. D. Bachmann, L. Balicas, M. K. Chan, B. J. Ramshaw, R. D. McDonald, F. F. Balakirev, M. Jaime, E. D. Bauer, and P. J. W. Moll, Nature {\bf 548}, 313-317 (2017).	
	




	
\bibitem{MBlume} M. Blume, Y. Y. Hsieh, J. Appl. Phys. {\bf 40}, 1249 (1969).
	
\bibitem{AFAndreev} A. F. Andreev and I. A. Grishchuk, Sov. Phys. JETP {\bf 60}, 267 (1984).	
	
\bibitem{KPenc} K. Penc and A. M. L\"{a}uchli, in {\it Introduction to Frustrated Magnetism}, edited by C. Lacroix, P. Mendels, and F. Mila (Springer, New York, 2011), p331.	

\bibitem{YKohama} Yoshimitsu Kohama, Hajime Ishikawa, Akira Matsuo, Koichi Kindo, Nic Shannon, and Zenji Hiroi, Proc. Natl. Acad. Sci. U.S.A. {\bf 116}, 10686-10690 (2019).

\bibitem{ROkazaki} R. Okazaki1, T. Shibauchi. H. J. Shi, Y. Haga, T. D. Matsuda, E. Yamamoto, Y. Onuki, H. Ikeda, and Y. Matsuda, Science {\bf 331}, 439-442 (2011).

\bibitem{EWRosenberg} Elliott W. Rosenberg, Jiun-Haw Chu, Jacob P. C. Ruff, Alexander T. Hristov, and Ian R. Fisher, Proc. Natl. Acad. Sci. U.S.A. {\bf 116}, 7232-7237 (2018).

\bibitem{CFang2008} Chen Fang, Hong Yao, Wei-Feng Tsai, JiangPing Hu, and Steven A. Kivelson, Phys. Rev. B {\bf 77}, 224509 (2008).

\bibitem{CXu2008} Cenke Xu, Markus M\"{u}ller, and Subir Sachdev, Phys. Rev. B {\bf 78}, 020501(R) (2008).	

\bibitem{RYu2015} Rong Yu and Qimiao Si, Phys. Rev. Lett. {\bf 115}, 116401 (2015).

\bibitem{GStewart_RMP} G. R. Stewart, Rev. Mod. Phys. {\bf 83}, 1589 (2011).

\bibitem{PDai_RMP} Pengcheng Dai, Rev. Mod. Phys. {\bf 87}, 855 (2015).

\bibitem{MTanatar2016} M. A. Tanatar, A. E. B\"{o}hmer, E. I. Timmons, M. Sch\"{u}tt, G. Drachuck, V. Taufour, K. Kothapalli, A. Kreyssig, S. L. Bud'ko, P. C. Canfield, R. M. Fernandes, and R. Prozorov, Phys. Rev. Lett. {\bf 117}, 127001 (2016).	

\bibitem{JHChu2} Jiun-Haw Chu, Hsueh-Hui Kuo, James G. Analytis, and Ian R. Fisher, Science {\bf 337}, 710-712 (2012).

\bibitem{WLv2009} Weicheng Lv, Jiansheng Wu, and Philip Phillips, Phys. Rev. B {\bf 80}, 224506 (2009).

\bibitem{CCLee2009} Chi-Cheng Lee, Wei-Guo Yin, and Wei Ku, Phys. Rev. Lett. {\bf 103}, 267001 (2009). 	
	
\bibitem{RMFernandes2012_1} R. M. Fernandes, A. V. Chubukov, J. Knolle, I. Eremin, and J. Schmalian, Phys. Rev. B {\bf 85}, 024534 (2012).

\bibitem{RMFernandes2012_2} Rafael M. Fernandes, and J\"{o}rg Schmalian, Supercond. Sci. Technol. {\bf 25}, 084005 (2012). 

\bibitem{SLiang2013} S. Liang, A. Moreo, and E. Dagotto, Phys. Rev. Lett. {\bf 111}, 047004 (2013).

\bibitem{RMFernandes2013} Rafael M. Fernandes, Anna E. B\"{o}hmer, Christoph Meingast, and J\"{o}rg Schmalian, Phys. Rev. Lett. {\bf 111}, 137001 (2013).

\bibitem{XLu2014} Xingye Lu, J. T. Park, Rui Zhang, Huiqian Luo, Andriy H. Nevidomskyy, Qimiao Si, and Pengcheng Dai, Science {\bf 345}, 657-660 (2014).

\bibitem{YSong2015} Yu Song, Xingye Lu, D. L. Abernathy, David W. Tam, J. L. Niedziela, Wei Tian, Huiqian Luo, Qimiao Si, and Pengcheng Dai, Phys. Rev. B {\bf 92}, 180504(R) (2015).

\bibitem{HZhai2009} Hui Zhai, Fa Wang, and Dung-Hai Lee, Phys. Rev. B {\bf 80}, 064517 (2009).

\bibitem{YGallais2013} Y. Gallais, R. M. Fernandes, I. Paul, L. Chauvi\`{e}re, Y.-X. Yang, M.-A. M\'{e}asson, M. Cazayous, A. Sacuto, D. Colson, and A. Forget, Phys. Rev. Lett. {\bf 111}, 267001 (2013).

\bibitem{MYi2011} Ming Yi, Donghui Lu, Jiun-Haw Chu, James G. Analytis, Adam P. Sorini, Alexander F. Kemper, Brian Moritz, Sung-Kwan Mo, Rob G. Moore, Makoto Hashimoto, Wei-Sheng Lee, Zahid Hussain, Thomas P. Devereaux, Ian R. Fisher, and Zhi-Xun Shen, Proc. Natl. Acad. Sci. U.S.A. {\bf 108}, 6878 (2011).

\bibitem{MHe2017} Mingquan He, Liran Wang, Felix Ahn, Fr\'{e}d\'{e}ric Hardy, Thomas Wolf, Peter Adelmann, J\"{o}rg Schmalian, Ilya Eremin, and Christoph Meingast, Nat. Commun. {\bf 8}, 804 (2017).

\bibitem{MHe2018} Mingquan He, Liran Wang, Fr\'{e}d\'{e}ric Hardy, Liping Xu, Thomas Wolf, Peter Adelmann, and Christoph Meingast, Phys. Rev. B {\bf 97}, 104107 (2018).

\bibitem{HLei2011} Hechang Lei, Emil S. Bozin, Kefeng Wang, and C. Petrovic, Phys. Rev. B {\bf 84}, 060506(R) (2011).

\bibitem{RAng2013} R. Ang, K. Nakayama, W.-G. Yin, T. Sato, Hechang Lei, C. Petrovic, and T. Takahashi, Phys. Rev. B {\bf 88}, 155102 (2013).

\bibitem{YSong2019} Yu Song, Huibo Cao, B. C. Chakoumakos, Yang Zhao, Aifeng Wang, Hechang Lei, C. Petrovic, and Robert J. Birgeneau, Phys. Rev. Lett. {\bf 122}, 087201 (2019).

\bibitem{QHuang} Q. Huang, Y. Qiu, Wei Bao, M. A. Green, J. W. Lynn, Y. C. Gasparovic, T. Wu, G. Wu, and X. H. Chen, Phys. Rev. Lett. {\bf 101}, 257003 (2008).

\bibitem{TYildirim2008} T. Yildirim, Phys. Rev. Lett. {\bf 101}, 057010 (2008). 

\bibitem{QMSi2008} Qimiao Si and Elihu Abrahams, Phys. Rev. Lett. {\bf 101}, 076401 (2008).

\bibitem{FKruger2009} Frank Kr\"{u}ger, Sanjeev Kumar, Jan Zaanen, and Jeroen van den Brink, Phys. Rev. B {\bf 79}, 054504 (2009). 

\bibitem{SI} See Supplemental Material for Curie-Weiss fits of the magnetic susceptibilities in the freestanding sample, and a discussion on measurements of the nematic susceptibility in the presence of strain.

\bibitem{CDhital2012} Chetan Dhital, Z. Yamani, Wei Tian, J. Zeretsky, A. S. Sefat, Ziqiang Wang, R. J. Birgeneau, and Stephen D. Wilson, Phys. Rev. Lett. {\bf 108}, 087001 (2012).

\bibitem{YSong2013} Yu Song, Scott V. Carr, Xingye Lu, Chenglin Zhang, Zachary C. Sims, N. F. Luttrell, Songxue Chi, Yang Zhao, Jeffrey W. Lynn, and Pengcheng Dai, Phys. Rev. B {\bf 87}, 184511 (2013).

\bibitem{XLu2016} Xingye Lu, Kuo-Feng Tseng, T. Keller, Wenliang Zhang, Ding Hu, Yu Song, Haoran Man, J. T. Park, Huiqian Luo, Shiliang Li, Andriy H. Nevidomskyy, and Pengcheng Dai, Phys. Rev. B {\bf 93}, 134519 (2016). 

\bibitem{RABorzi2007} R. A. Borzi, S. A. Grigera, J. Farrell, R. S. Perry, S. J. S. Lister, S. L. Lee, D. A. Tennant, Y. Maeno, and A. P. Mackenzie, Science {\bf 315}, 214-217 (2007).

\bibitem{ZLiu2016} Zhaoyu Liu, Yanhong Gu, Wei Zhang, Dongliang Gong, Wenliang Zhang, Tao Xie, Xingye Lu, Xiaoyan Ma, Xiaotian Zhang, Rui Zhang, Jun Zhu, Cong Ren, Lei Shan, Xianggang Qiu, Pengcheng Dai, Yi-feng Yang, Huiqian Luo, and Shiliang Li, Phys. Rev. Lett. {\bf 117}, 157002 (2016).

\bibitem{HHKuo2013} Hsueh-Hui Kuo, Maxwell C. Shapiro, Scott C. Riggs, and Ian R. Fisher, Phys. Rev. B {\bf 88}, 085113 (2013).

\bibitem{JLi2016} J. Li, D. Zhao, Y. P. Wu, S. J. Li, D. W. Song, L. X. Zheng, N. Z. Wang, X. G. Luo, Z. Sun, T. Wu, and X. H. Chen, arXiv: 1611.04694 (2016).

\bibitem{XLiu2019} Xi Liu, Ran Tao, Mingqiang Ren, Wei Chen, Qi Yao, Thomas Wolf, Yajun Yan, Tong Zhang, and Donglai Feng, Nat. Commun. {\bf 10}, 1039 (2019).

\bibitem{KIshida2018} K. Ishida, M. Tsujii, S. Hosoi, Y. Mizukami, S. Ishida, A. Iyo, H. Eisaki, T. Wolf, K. Grube, H. v. L\"{o}hneysen, R. M. Fernandes, and T. Shibauchi, arXiv: 1812.05267 (2018).

\bibitem{XFWang2009} X. F. Wang, T. Wu, G. Wu, H. Chen, Y. L. Xie, J. J. Ying, Y. J. Yan, R. H. Liu, and X. H. Chen, Phys. Rev. Lett. {\bf 102}, 117005 (2009).

\bibitem{JYang2010} Jinhu Yang, Mami Matsui, Masatomo Kawa, Hiroto Ohta, Chishiro Michioka, Chiheng Dong, Hangdong Wang, Huiqiu Yuan, Minghu Fang, and Kazuyoshi Yoshimura, J. Phys. Soc. Jpn. {\bf 79}, 074704 (2010).

\bibitem{WZhang2016} Wenliang Zhang, J. T. Park, Xingye Lu, Yuan Wei, Xiaoyan Ma, Lijie Hao, Pengcheng Dai, Zi Yang Meng, Yi-feng Yang, Huiqian Luo, and Shiliang Li, Phys. Rev. Lett. {\bf 117}, 227003 (2016).

\bibitem{RMFernandes2019} Rafael M. Fernandes, Peter P. Orth, and J\"{o}rg Schmalian, Annu. Rev. Condens. Matter Phys. {\bf 10}, 133-154 (2019).

\bibitem{KK} K. I. Kugel and D. I. Khomskii, Sov. Phys. Usp. {\bf 25}, 231 (1982).

\bibitem{Oles2012} A. M. Ole\'{s}, J. Phys.: Condens. Matter {\bf 24}, 313201 (2012).

\bibitem{NQureshi2012} N. Qureshi, P. Steffens, S. Wurmehl, S. Aswartham, B. B\"{u}chner, and M. Braden, Phys. Rev. B {\bf 86}, 060410(R) (2012).

\bibitem{YSong2013_2} Yu Song, Louis-Pierre Regnault, Chenglin Zhang, Guotai Tan, Scott V. Carr, Songxue Chi, A. D. Christianson, Tao Xiang, and Pengcheng Dai, Phys. Rev. B {\bf 88}, 134512 (2013).



\end{thebibliography}
\end{document}